\documentclass[11pt]{article}

\usepackage{amsmath}

\usepackage[cp866]{inputenc}

\newcommand{\N}{N\raise.7ex\hbox{\underline{$\circ $}}$\;$}

\begin{document}

\begin{center}

{\bf  E.M. Ovsiyuk,   V.V. Kisel,  V.M. Red'kov  \\[3mm]
On exact solutions of the  Dirac equation \\ in a homogeneous
magnetic field  in the Riemann spherical  space}
\\[3mm]
{\em Institute of Physics, NAS of Belarus\\
Belarussian  State Pedagogical University}

\end{center}

\begin{quotation}

There are constructed exact solutions of the quantum-mechanical Dirac equation for
a spin S=1/2 particle in the space of constant positive curvature, spherical Riemann space,
in presence of an external magnetic field, analogue of the homogeneous magnetic field in the
 Minkowski  space. A generalized formula for energy levels, describing quantization of the
  motion of the particle in magnetic field on the background of the Riemann space geometry, is obtained.

\end{quotation}

{\bf  1.  Introduction }

\vspace{3mm}

The  quantization  of a quantum-mechanical particle in the
homogeneous magnetic field belongs to classical  problems in
physics  \cite{1,2,3}. In  \cite{4,5,6}, exact solutions for  a
scalar particle in extended problem, particle in  external
magnetic field on the background of Lobachevsky $H_{3}$  and
Riemann $S_{3}$ spatial geometries were found. A corresponding
system in the frames of classical mechanics was examined in
 \cite{7,8,9}. In the present paper, we consider a similar problem for a particle with spin $1/2$ described
 by Dirac equation in Lobachevsky space in presence of the external magnetic field.

\vspace{2mm}

{ \bf 2. Cylindric coordinates ant the Dirac equation in spherical
 space $H_{3}$ $S_{3}$}

\vspace{2mm}

In the spherical Riemann  space $S_{3}$, let us use an extended cylindric coordinates
\begin{eqnarray}
dS^{2} =  dt^{2} -   \cos^{2} z ( d r^{2} + \sin^{2} r \; d
\phi^{2} ) + dz^{2}\; ]\; , \nonumber
\\
 z \in [-\pi /2 , + \pi /2
]\; , \qquad r \in [0, + \pi ] , \qquad \phi \in [0, 2 \pi ] \; ,
\nonumber
\\
u^{1} = \cos z  \sin r \cos \phi \; , \;  u^{2} = \cos z \;
\sin r \sin \phi \; ,\;
u^{3} = \sin z \; , \; u^{0} = \cos z  \cos r \; ;
\label{2.4}
\end{eqnarray}

\noindent the curvature radius $\rho$ is taken as a  unit  of the
length. An analogue of usual homogeneous magnetic field is defined
as  \cite{4,5,6}
\begin{eqnarray}
 A_{\phi} = -2B \sin^{2} {r \over 2} = B\; ( \cos r -1 )\; .
\label{2.5}
\end{eqnarray}

To coordinates  (\ref{2.4})  there corresponds the tetrad
\begin{eqnarray}
 e_{(a)}^{\beta}(x) = \left |
\begin{array}{llll}
1 & 0 & 0 & 0 \\
0 & \cos^{-1}z & 0 & 0 \\
0 & 0 & \cos^{-1}z\;\sin^{-1} r & 0 \\
0 & 0 & 0 & 1
\end{array} \right | \; .
\label{2.6}
\end{eqnarray}

 \noindent Christoffel symbols  $\Gamma^{r}_{\;\;jk }$ and Rici rotation coefficients
 $\gamma_{abc}$  are
\begin{eqnarray}
\Gamma^{r}_{\;\;jk } = \left | \begin{array}{ccc}
0 & 0 & -\mbox{tg}\;z \\
0 & - \sin r \cos r & 0 \\
- \mbox{tg}\;z & 0 & 0
\end{array} \right |  , \qquad
\Gamma^{\phi}_{\;\;jk } = \left | \begin{array}{ccc}
0 & \mbox{ctg}\; r & 0\\
\mbox{ctg}\; r & 0 &- \mbox{tg}\; z \\
0 & -\mbox{tg}\; z & 0
\end{array} \right |  ,
\nonumber
\end{eqnarray}
\begin{eqnarray}
 \Gamma^{z}_{\;\;jk } = \left | \begin{array}{ccc}
\sin z \cos z & 0 & 0\\
0 & \sin z \; \cos z \sin^{2} r & 0 \\
0 & 0 & 0
\end{array} \right | ,
\nonumber
\end{eqnarray}
\begin{eqnarray}
\gamma_{12 2} =
 { 1 \over \cos z \mbox{tan}\; r} \; , \;
 \gamma_{31 1} =
 -\mbox{tan}\; z\; , \; \gamma_{32 2} =
 -\mbox{tan}\; z\; .
 \nonumber
 \label{2.8}
\end{eqnarray}

A general covariant Dirac equation (for more detail see [10])  takes the form
\begin{eqnarray}
\left [ i \gamma^{0} \partial_{t}  + {i \gamma^{1}  \over  \cos z}
 (    \partial_{r} + {1 \over 2}  {1 \over \mbox{tan}\;  r}  )
 + \gamma^{2}  {    i   \partial_{\phi} - e  B (\cos r -1)    \over \cos z \sin r}    +
 i \gamma^{3} ( \partial_{z} - \mbox{tan}\; z    ) - M \right  ]  \Psi = 0\; .
\label{2.11}
\end{eqnarray}

With the  substitution $ \Psi = \varphi/  ( \sqrt{\sin r}\;  \cos
) $ eq. (\ref{2.11}) becomes simpler
\begin{eqnarray}
\left [  i \gamma^{1}  {\partial \over \partial r }  +  \gamma^{2}
\;
 {i \partial_{\phi} - e B (\cos r -1)  \over \sin r }   +
  \cos z \left  (i  \gamma^{0} {\partial  \over  \partial t}  +
  i  \gamma^{3} { \partial \over  \partial z} -  M  \right  )  \right ] \varphi =0 \; .
\label{3.9}
\end{eqnarray}

Solutions of this equation will be searched in the form
\begin{eqnarray}
\varphi = e^{-i\epsilon t} e^{im \phi} \left | \begin{array}{c}
f_{1}(r,z)\\ f_{2}(r,z)\\
f_{3}(r,z)\\
f_{4}(r,z) \
\end{array} \right | ;
\nonumber
\end{eqnarray}

\noindent so that
\begin{eqnarray}
\left [  i \gamma^{1}  {\partial \over \partial r }  -
\mu (r) \gamma^{2}   +
  \cos z  \left  ( \epsilon   \gamma^{0}   +
  i  \gamma^{3} { \partial \over  \partial z} -  M  \right  )  \right ]  \left | \begin{array}{c}
f_{1}(r,z)\\ f_{2}(r,z)\\
f_{3}(r,z)\\
f_{4}(r,z) \
\end{array} \right |
 =0 \; ,
 \label{3.10}
\end{eqnarray}

\noindent
 where
$ \mu (r) = [m -  e B (\mbox {ch}\,  r -1) ]/ \;  \mbox{sh}\,  r $.
Taking the  Dirac matrices in spinor basis,
we get radial equations for  $f_{a}(t,z)$
\begin{eqnarray}
  ( {\partial \over \partial r }  + \; \mu  ) \; f_{4}
   + \cos z  \;  { \partial f_{3} \over  \partial z}
+  i \; \cos z  \; ( \epsilon   f_{3}   -  M  f_{1} ) =0\,,
\nonumber
\\
( {\partial \over \partial r }  - \; \mu  ) \;f_{3} -  \cos z  \;
{ \partial f_{4} \over  \partial z}
 +   i  \; \cos z \; (  \epsilon  f_{4}   - M  f_{2} ) =0\,,
\nonumber
\\
 ( {\partial \over \partial r }  +   \; \mu )  \; f_{2}
+ \cos z \;  { \partial f_{1} \over  \partial z}
 - i   \; \cos z  \; ( \epsilon f_{1} -M f_{3} ) =0\,,
\nonumber
\\
  ( {\partial \over \partial r }  -  \; \mu  ) \; f_{1}
-  \cos z  \; { \partial f_{2} \over  \partial z} - i  \; \cos z
\; (   \epsilon f_{2} -M f_{4} ) =0\, .
\label{3.15}
\end{eqnarray}

\noindent
With linear restriction
$
f_{3} = A f_{1} ,  \; f_{4} = A f_{2}$, where
\begin{eqnarray}
\epsilon -{M \over A} = - \epsilon + M A \qquad  \Longrightarrow
\qquad A = A_{1,2}= {\epsilon \pm p \over M} \; ,\qquad  ( p = +
\sqrt{\epsilon^{2} - M^{2}}  )  \label{3.18}
\end{eqnarray}

\noindent
eqs. (\ref{3.15}) give
\begin{eqnarray}
 ( {\partial \over \partial r }  +   \; \mu )  \; f_{2}
+ \cos z \;  { \partial f_{1} \over  \partial z}
 + i   \; \cos z  \; ( - \epsilon  + M  A  ) \;  f_{1} =0\,,
\nonumber
\\
  ( {\partial \over \partial r }  -  \; \mu  ) \; f_{1}
-  \cos z  \; { \partial f_{2} \over  \partial z} + i  \; \cos z
\; (  -  \epsilon   + M  A ) \;  f_{2} =0\, .
\label{3.19}
\end{eqnarray}

Thus, we have two possibilities

\vspace{4mm}
 $
 A =   (\epsilon + p)/M  \; ,
 $
\begin{eqnarray}
 ( {\partial \over \partial r }  +   \; \mu )  \; f_{2}
+ \cos z \;  ( { \partial  \over  \partial z}
 +  i  p   \; )\;   f_{1} =0\,,\qquad
  ( {\partial \over \partial r }  -  \; \mu  ) \; f_{1}
-  \cos z  \; ({ \partial  \over  \partial z} - i  \; p\; )\;
f_{2} =0\, ;
\label{3.20a}
\end{eqnarray}

$
 A =   (\epsilon - p)/M  \; ,
$
\begin{eqnarray}
 ( {\partial \over \partial r }  +   \; \mu )  \; f_{2}
+ \cos z \;  ( { \partial  \over  \partial z}
 -  i  p\;)   \;    f_{1} =0\,,\qquad
  ( {\partial \over \partial r }  -  \; \mu  ) \; f_{1}
-  \cos z  \; ( { \partial  \over  \partial z} + i  \; p\; ) \;
f_{2} =0\, .
\label{3.20b}
\end{eqnarray}

For definiteness, let us consider the system   (\ref{3.20a}) (transition to the case
 (\ref{3.20b}) is performed by the formal change $p \Longrightarrow -p$).
 Let us search solutions in the form
\begin{eqnarray}
f_{1} = Z_{1} (z) \; R_{1} (r) \; ,  \qquad  f_{2} = Z_{2}(z)
\; R_{2} (r)  \; .
\nonumber
\end{eqnarray}

\noindent Introducing the a separating constant $\lambda$
\begin{eqnarray}
 \cos z  \; ({ d  \over d z} + i  \; p\; )\;
Z_{1}  = \lambda  \; Z_{2} \; , \qquad
\cos z  \; ({ d  \over  d z} - i  \; p\; )\; Z_{2} = \lambda \;
Z_{1}  \; ,
\label{3.23}
\end{eqnarray}

\noindent  we get the radial system
\begin{eqnarray}
 ( {d \over d r }  +   \; \mu )  \;  R_{2} +
\lambda  \; R_{1}  =0\,, \qquad
   ( { d \over d r }  -  \; \mu  ) \;  R_{1}
-  \lambda   \; R_{2}  =0\, .
\label{3.24}
\end{eqnarray}

{\bf 3. Solution of the equation in  $z$-variable}

\vspace{2mm}

From  (\ref{3.23}) it follows the second-order differential equation for $Z_{1}(z)$
\begin{eqnarray}
\left ( {d^{2} \over dz} -
{\sin z \over \cos z} {d  \over
dz}+   p^{2}-ip{\sin z\over \cos z}-{\lambda^{2}\over
\cos^{2}z} \; \right  )  Z_{1}=0\,.
\label{3.25}
\end{eqnarray}

\noindent In a new variable  $
 y = (1 +  i\mbox{tg}\; z ) /  2)$, eq. (\ref{3.25}) gives
 \begin{eqnarray}
\left [ 4y (1-y)
 {d ^{2}\over d y^{2}} +
2  (1-2y) {d \over d y}  -
     p^{2} ( {1 \over 1- y}  + {1 \over y} ) + p (  {1 \over 1-y } - {1 \over y}  )
+  4 \lambda^{2}
 \right ] \; Z_{1}=0 \, .
\label{3.27}
\end{eqnarray}

\noindent
With substitution
$
Z_{1} = y^{A} (1-y)^{C} Z (y)$ we arrive at the equation
\begin{eqnarray}
4y\,(1-y)\,{d^{2}Z\over dz^{2}}+4\left[2A\,+\,{1\over
2}-(2A\,+\,2C\,+1)\,y\right]{dZ\over dz}+
\nonumber
\\
+\left[{2A\,(2A-1)-p\,(p+1)\over y}+{2C\,(2C-1)-p\,(p-1)\over
1-y}-4\,(A+C)^{2}+4\lambda^{2}\right]Z=0\,.
\nonumber
\end{eqnarray}

\noindent Requiring
\begin{eqnarray}
 A= -{p\over
2}\,,\; {p+1\over 2} \, , \qquad C = {p\over 2}
\,,\;{1-p\over 2} \, ,
\label{3.29}
\end{eqnarray}

\noindent for  $Z_{1}$  we get an equation of hypergeometric type
\begin{eqnarray}
y\,(1-y)\,{d^{2}Z\over dz^{2}}+\left[2A\,+\,{1\over
2}-(2A\,+\,2C\,+1)\,y\right]{dZ\over
dz}-\left[(A+C)^{2}-\lambda^{2}\right]Z=0\, ,
\nonumber
\end{eqnarray}

\noindent
where
\begin{eqnarray}
\alpha = \lambda+A+C \;, \qquad \beta = -\lambda+A+C \;, \qquad
\gamma = 2A\,+\,{1\over 2} \; ,
\nonumber
\\
Z_{1} =  \left ( {e^{iz} \over \cos z} \right  )^{A} \; \left  (
{e^{-iz} \over \cos z} \right )^{C} F(\alpha,  \; \beta,  \;
\gamma ; \;    {e^{iz} \over 2 \cos z} ) \; .
\label{3.32}
\end{eqnarray}

There arise four   possibility depending on the choice of
 $A$ and $C$. For definiteness let us suppose that $\lambda
>0$ which does not affect generality of consideration:

\begin{eqnarray}
\underline{\mbox{Variant }\; 1}\;, \qquad A = {p+1 \over 2} \;,
\qquad C = {1 - p \over 2} \;, \qquad A+C = 1 \;, \qquad A - C =p
\; ,
\nonumber
\\
\beta =  - N \; , \qquad   \Longrightarrow  \qquad
\underline{\mbox{spectrum}} \qquad \lambda = 1 +  N \; , \qquad N =
0, 1,2, ...
\nonumber
\\
 \alpha = \lambda+1, \qquad \gamma = p+{3\over 2}\,, \qquad
Z_{1} = {e^{ipz}  \over  \cos z  } \; F (\lambda+1,- N \;
,p+{3\over 2}\,; {e^{iz} \over 2 \cos z}).
\label{3.34}
\end{eqnarray}

\begin{eqnarray}
\underline{\mbox{Variant  }\; 2}\; , \qquad A = -{p \over 2} \;,
\qquad C = {p \over 2} \;, \qquad A+C =  0 \;, \;\; A- C = -p  \;
,
\nonumber
\\
\beta =  - N \; , \qquad   \Longrightarrow  \qquad
\underline{\mbox{spectrum}} \qquad
  \lambda = N  \; , \qquad  N = 0,1,2, ... ,
\nonumber
\\
 \alpha = + N \; , \qquad \gamma = - p+{1\over 2}\,, \qquad
Z_{1} = e^{-ipz}   \; F (\lambda,- N \; ,-p+{1\over 2}\,;{e^{iz}
\over 2 \cos z}).
\label{3.35}
\end{eqnarray}

In these two cases, we do not have found quantization for $p$, instead we have some quantization rules
for  $\lambda$. However,   quite different quantization rules (as expected)
 will  follow from consideration of the differential equation in $r$-variable.
 By this reason, these cases 1 and 2 will not be considered  in the following.

\begin{eqnarray}
\underline{\mbox{Variant  }\; 3,} \qquad A = {p+1 \over 2} \;,
\qquad C = {p \over 2} \;, \qquad A+C = p +1/2\;, \qquad A- C =1/2
\; ,
\nonumber
\\
\beta =  - N \; \;    \Longrightarrow  \;\;
\underline{\mbox{spectrum}} \qquad p =+\sqrt{\epsilon^{2}- M^{2}} =
\lambda - (N  +1/2 ) \; , \qquad N = 0,1,2, ..., N_{max}
\nonumber
\\
 \alpha = \lambda+p+{1\over 2} \; , \qquad \gamma = p+{3\over2} \; ,\;
Z_{1} = {e^{iz/2}  \over ( \cos z  ) ^{p+1/2} } \; F
(\lambda+p+{1\over 2},  \; - N,
 \;p+{3\over2} ; \;    {e^{iz} \over 2 \cos z}) \; .
\label{3.33}
\end{eqnarray}

\begin{eqnarray}
\underline{\mbox{ Variant } \; 4},  \qquad A =  -{p \over 2} \;,
\qquad C = {1 - p \over 2} \;, \qquad A+C = -p +1/2  \;, \;\; A -
C = -1/2  \; ,
\nonumber
\\
\alpha  =  - N \; , \qquad   \Longrightarrow  \qquad
\underline{\mbox{spectrum}} \qquad
  p = \sqrt{\epsilon^{2} - M^{2}}  = \lambda +  (N + 1/2)  \; ,
\nonumber
\\
 \beta = - \lambda - p + {1\over 2} \; , \;  \gamma = -p + {1\over 2}\,,
 \;
Z_{1} = {e^{-iz/2}  \over  (\cos z )^{-p +1/2} } \;
 F (\lambda-p+{1\over 2},- N \; , -p+{1\over 2}\,;{e^{iz} \over 2 \cos z}) \; .
\label{3.36}
\end{eqnarray}

\noindent
To the cases 3 and 4 there correspond different energy spectra:
\begin{eqnarray}
 3\; , \qquad p
=+\sqrt{\epsilon^{2}-m^{2}} = \lambda - (N  +1/2 ) \;  ;
\nonumber
\\
 4 \; ,   \qquad   p
=+\sqrt{\epsilon^{2}-m^{2}} = \lambda + (N  +1/2 ) \; .
\label{3.37}
\end{eqnarray}

\vspace{2mm}

{\bf 4. Solution of the equations in  $r$-variable}

\vspace{2mm}

From eqs. (\ref{3.24}) it follows a second-order differential equation for $R_{1}$
(for brevity let  $eB$  be  noted as $B$)
\begin{eqnarray}
{d^{2}R_{1}\over dr^{2}}+\left[{m\,\cos  r-B\,(\cos  r-1)\over
\sin^{2}  r }-{[m+B\,(\cos  r-1)]^{2}\over \sin^{2}  r
}+\lambda^{2}\right]R_{1}=0\,.
\label{3.35}
\end{eqnarray}

\noindent With  a new variable $ y= (1+ \cos r )/ 2 $,  eq. (\ref{3.35}) reads
\begin{eqnarray}
y(1-y){d^{2}R_{1}\over dy^{2}}+\left({1\over
2}-y\right){dR_{1}\over dy}-
\nonumber
\\
-\left[-\lambda^{2}+{m^{2}\over 4} \left({1\over y}+{1\over
1-y}\right)+{m\over 4} \left({1\over y}-{1\over 1-y}\right)-{mB
\over y}-B^{2} \left(1-{1\over y}\right)-{B\over
2y}\right]R_{1}=0\,.
\label{3.35}
\end{eqnarray}

\noindent
With the substitution
$ R_{1} = y^{A} (1-y)^{C} R (y)$, we get
\begin{eqnarray}
y(1-y){d^{2}R\over dy^{2}}+ \left [  2A + {1\over 2}-(2A+2C +1)\;
y  \right] \,{dR\over dy}+
\nonumber
\\
+\left [ { A^{2}-A/2-m^{2}/4-m/4+ mB - B^{2}+B/2\over y}+\right.
\nonumber
\\
\left.+{C^{2}-C/2- m^{2}/4 + m/4\over 1-y}
-(A+C)^{2}+\lambda^{2}+B^{2}\right] R=0\, .
\nonumber
\end{eqnarray}

\noindent Requiring
\begin{eqnarray}
 A= {2B - m\over 2}\; ,\;  {-2B + m +1 \over 2} \, , \qquad
 C = {m\over 2} ,\; {1 -m \over 2} \, ,
 \label{3.37}
 \end{eqnarray}

\noindent we arrive at a differential equation of hypergeometric type
\begin{eqnarray}
y(1-y){d^{2}R\over dy^{2}}+ \left [  2A + {1\over 2}-(2A+2C +1)\;
y  \right] \,{dR\over dy} -\left[
(A+C)^{2}-\lambda^{2}-B^{2}\right]R=0\, .
\label{3.38}
\end{eqnarray}

\noindent
where
\begin{eqnarray}
\alpha = A+C - \sqrt{B^{2}+\lambda^{2}}\;, \qquad \beta = A+C
+\sqrt{B^{2}+\lambda^{2}} \;, \qquad \gamma = 2A\,+\,{1\over 2} \; ,
\nonumber
\\
R_{1} = ( 1 +  \cos  r) ^{A} \; (1 - \cos  r)^{C} \;
F(\alpha, \beta, \gamma ;\; {1 +  \cos  r \over 2} ) \; .
 \label{3.39}
 \end{eqnarray}

In order to have a finite  solution at the origin $r=0$ (that corresponds to  the half-curve
 $u_{0} = + \cos z,
\;  u_{3} = \sin  z  ,  u_{1}=0 ,  u_{2}=0$) and at  $r = \pi$
(that corresponds to the other part of the curve,  $
u_{0} = - \cos z  ,   u_{3} = \sin  z  ,  u_{1}=0 ,
u_{2}=0$ ),  we  must take positive values for
$A$ and  $C$:
\begin{eqnarray}
R_{1} = ( 1 + \cos r) ^{A} \; (1 - \cos  r)^{C} \;  F(\alpha,
\beta, \gamma ;\; {1 +   \cos  r \over 2} ) \; , \qquad A >
0 \; , \;  \; C > 0 \; .
\label{3.41}
\end{eqnarray}

Let us consider all the four variants  (for definiteness let us assume that  $B>0$ ):
\begin{eqnarray}
\mbox{Variant } \; 1\; , \qquad  \qquad C = {m \over 2}  > 0 , \qquad A =
 {2B - m  \over 2}  > 0  \; ,\qquad  C+A = + B    \; .
\nonumber
\\
\mbox{Variant } \; 2 \; , \qquad C = {m \over 2}  > 0 , \qquad A =
{ m +1 - 2B \over 2} > 0 \; , \qquad  C+A = -B + m + {1 \over 2}  \; .
\nonumber
\\
\mbox{Variant } \; 3\; ,  \qquad C = {1-m \over 2}  > 0 , \qquad A
= { m +1 - 2B \over 2} >  0    \; ,\qquad  C+A = 1-B   > 0  \; .
\nonumber
\\
\mbox{Variant } \; 4\; , \qquad C = {1-m \over 2}  > 0 , \qquad A =
 {2B - m  \over 2} >  0 \; ,\qquad  C+A =  B  - m  +{1 \over 2}
\; .
\label{3.42}
\end{eqnarray}

To these there correspond the following solutions:

\vspace{2mm}

\underline{$ \mbox{Variant  } \; 1\; , \qquad 0 < m < 2B \; ,
$}
\begin{eqnarray}
R_{1} = ( 1 +  \cos  r) ^{(2B-m)/2} \; (1 -  \cos r)^{m/2} \;  F(\alpha, \beta, \gamma ;\; {1 +  \cos  r \over 2} )
\; ,
\nonumber
\\
\alpha = - n\;, \qquad  \beta = 2B + n \;, \qquad  \gamma =2B - m -1 \; ,
\nonumber
\\
 \underline{\mbox{spectrum}} \qquad \qquad  \lambda^{2} = 2Bn + n^{2}\; , \qquad n = 0, 1, 2,\;  ...
\label{3.43}
\end{eqnarray}

\vspace{5mm}

\underline{$ \mbox{Variant } \; 2\; , \qquad m > 0 \; , \;\; m > 2B -1 \;$ } ,
\begin{eqnarray}
R_{1} = ( 1 + \cos r) ^{(m+1 - 2B)/2 } \; (1 - \cos
r)^{m/2} \;  F(\alpha, \beta, \gamma ;\; {1 +  \cos  r \over 2} )
\; ,
\nonumber
\\
\alpha = -n \; , \qquad    n +  m +1/2 - B = \sqrt{B^{2}+\lambda^{2}}\;,
\nonumber
\\
\beta = n + 2m +1 - 2B  \;, \qquad \gamma = - 2B + m +3/2 \; .
\nonumber
\\
\alpha = -n \;  \qquad \Longrightarrow \qquad \sqrt{B^{2} +
\lambda^{2}} = B + (m + n +{1\over 2 } ) \; ,
\nonumber
\\
\mbox{spectrum}\qquad \lambda^{2} = -2B \; (m + n +{1\over 2 }) +  (n + m  +{1\over 2 })^{2}  \; ,
\nonumber
\\
\lambda^{2} > 0 \qquad  \Longrightarrow \qquad
n + m +{1 \over 2} > 2B \; .
\label{3.44}
\end{eqnarray}

\vspace{5mm}

\underline{$ \mbox{Variant } \; 3\; , \qquad  m < 1\;, \; m > 2B -1 \;, \; 0<B<1 $ ,}
\begin{eqnarray}
R_{1} = ( 1 +  \cos  r) ^{(m+1 -2B)/2} \; (1 -  \cos r)^{(1-m)/2} \;  F(\alpha, \beta, \gamma ;\; {1 +  \cos  r \over
2} ) \; ;
\nonumber
\\
\alpha =  -n \; , \qquad  1 - B  + n =  \sqrt{B^{2}+\lambda^{2}}\;,
\nonumber
\\
 \beta = 2 -2B + n \;, \qquad \gamma = -2B + m +3/2  \; ,
\nonumber
\\
\mbox{spectrum}\qquad \lambda^{2} = - 2B \; (n + 1) +  ( n
+ 1 )^{2} \;, \;\; n+1 > 2B \;  .
\label{3.45}
\end{eqnarray}

\vspace{5mm}

\underline{$ \mbox{Variant } \; 4\; , \qquad   m <  1  \; ,\; m < 2B \; , \, m < B +1/2 $}
\begin{eqnarray}
R_{1} = ( 1 + \cos  r) ^{(2B-m)/2}  \; (1 -  \cos r)^{(1-m)/2} \;  F(\alpha, \beta, \gamma ;\; {1 +  \cos  r \over
2} ) \; ,
\nonumber
\\
\alpha = -n\;, \qquad  B + n  - m + {1 \over 2}  =  \sqrt{B^{2}+\lambda^{2}}\;,
\nonumber
\\
 \beta = 2B - 2m + 1 + n  \;, \qquad \gamma = 2B  - m + {1\over 2}
\; ,
\nonumber
\\
\mbox{spectrum} \qquad \lambda^{2} = 2B \; ( n   -m  +{1\over 2 } ) + ( n - m   + {1\over 2
})^{2} \; .
 \label{3.44}
 \end{eqnarray}

In the end there should be given two  clarifying additions.
In fact, the above used  relationship
$-i \partial_{\phi } \Psi = m  \; \Psi $ represents transformed from cartesian coordinates to cylindrical
an eigen-value  equation for the third projection of the the total angular momentum of the Dirac particle
\begin{eqnarray}
\hat{J}_{3} \Psi_{Cart}  = ( -i {\partial \over  \partial \phi }
  + \Sigma_{3} )\; \Psi_{Cart} = m  \; \Psi = m\; \Psi_{Cart} \; ;
\label{3.45}
\end{eqnarray}

\noindent  this means that for  the quantum number
 $m$ are permitted half-integer values $ m = \pm {1 \over 2}, \; \pm {3\over 2}, ... $

When using ordinary units (in the system SI), we should change the symbol
 $B$ into $ eB \rho^{2}  /  \hbar $. Therefore, in the  limit of vanishing curvature
 $\rho \rightarrow  + \infty$ , from  the four noted classes of solutions
only two of them remain:
\begin{eqnarray}
1. \qquad \rho \rightarrow  + \infty \qquad m = +1/2, +3/2, ...\; ; \qquad
4. \qquad \rho \rightarrow  + \infty \qquad m = -1/2, -3/2, ...\; .
\nonumber
\end{eqnarray}

\noindent they coincide with known results for the energy spectrum for particle  with spin 1/2 in Minkowski space.

Authors are  grateful to participants of the Scientific Seminar of
the Laboratory of theoretical physics, Institute of physics of
National Academy od Sciences of Belarus, for discussion.

\end{document}